\documentclass[10pt,aps,amsmath,showpacs,amssymb,nofootinbib,pre,twocolumn,epsfig]{revtex4}
\usepackage{bmpsize}
\usepackage[dvips,usenames]{color}
\usepackage{amsmath}
\newcommand{\be}{\begin{equation}}
\newcommand{\ee}{\end{equation}}
\newcommand{\av}[1]{\langle {#1} \rangle}

\usepackage{graphicx}
\usepackage{lipsum}

\begin{document}

\title{What makes us humans: Differences in the critical dynamics 
underlying the human and fruit-fly connectome}

\author{G\'eza \'Odor (1), Gustavo Deco (2) and Jeffrey Kelling (3)}
\affiliation{(1) Institute of Technical Physics and Materials Science,
Center for Energy Research, P. O. Box 49, H-1525 Budapest, Hungary \\ 
(2) Center for Brain and Cognition, Theoretical and Computational Group,
Universitat Pompeu Fabra / ICREA, Barcelona, Spain \\
(3) Department of Information Services and Computing,
Helmholtz-Zentrum Dresden - Rossendorf, P.O.Box 51 01 19, 01314 Dresden, Germany
}


\date{\today}


\begin{abstract}


Previous simulation studies on human connectomes suggested, that critical
dynamics emerge subcrititcally in the so called Griffiths Phases.
Now we investigate this on the largest available brain network, the 
$21.662$ node fruit-fly connectome, using the Kuramoto synchronization
model. As this graph is less heterogeneous, lacking modular structure
and exhibit high topological dimension, we expect a difference 
from the previous results.
Indeed, the synchronization transition is mean-field like,
and the width of the transition region is larger than in
random graphs, but much smaller than as for the KKI-18 human connectome. 
This demonstrates the effect of modular structure and dimension 
on the dynamics, providing a basis for better understanding the 
complex critical dynamics of humans.

\end{abstract}

\maketitle

\section{Introduction}

Power law (PL) distributed neuronal avalanches were shown in neuronal 
recordings (spiking activity and local field potentials, LFPs) of neural 
cultures in vitro~\cite{BP03,Mazzoni-2007,Pasquale-2008,Fried}, LFP 
signals in vivo~\cite{Hahn-2010}, field potentials and functional magnetic
resonance imaging (fMRI) blood-oxygen-level-dependent
(BOLD) signals in vivo~\cite{Shriki-2013,Tagliazucchi-2012}, voltage
imaging in vivo~\cite{Scott-2014}, and $10$--$100$ single-unit or
multi-unit spiking and calcium-imaging activity in
vivo~\cite{Pris,Bellay-2015,Hahn-2017,Seshadri-2018}.
Furthermore, source reconstructed magneto- and electroencephalographic
recordings (MEG and EEG), characterizing the dynamics of ongoing cortical activity, have also shown robust PL scaling in neuronal long-range temporal
correlations. These are at time scales from seconds to hundreds 
of seconds and describe behavioral scaling laws consistent with concurrent
neuronal avalanches~\cite{brainexp}. However, the measured scaling exponents
do not seem to be universal.

Besides the experiments theoretical research provide evidence that
the brain operates in a critical state between sustained
activity and an inactive
phase~\cite{ChialvBak1999_LearningMistakes, RChial2004_CriticalBrainNetworks,
Chialv2006_OurSensesCritical,Chialv2007_BrainNear,ChialvBalenzFraima2008_Brain:What,
FraimaBalenzFossChialv2009_Ising-likeDynamicsLarge-scale,
expert_self-similar_2011,FraimaChialv2012_WhatKindNoise,
Deco12,66,Senden16,MArep}.

Criticality in general occurs at continuous, second order phase transitions
and an ubiquitous phenomenon in nature as systems can benefit
many ways from it. As correlations and fluctuations diverge~\cite{Hai}
in neural systems working memory and long-range interactions can be generated 
spontaneously~\cite{Johnson} and the sensitivity to external signals is maximal. 
Furthermore, it has also been shown that information-processing capabilities 
are optimal near the critical point. 
Therefore, systems tune themselves close to criticality via self-organization 
(SOC)~\cite{SOC,Chi10}, presumably slightly below to avoid blowing over 
excitation. However, criticality is not a necessary condition for 
power-law statistics to appear, see~\cite{PhysRevE.95.012413}, so the
presented numerical results do not provide a full proof for the
criticality hypothesis of the whole brain, but remain within the 
validity of model assumptions. 

Besides, if quenched heterogeneity (that is called disorder compared
to homogeneous system) is present, rare-region (RR) effects~\cite{Vojta2006b}
and an extended semi-critical region, known as Griffiths Phase 
(GP)~\cite{Griffiths} can emerge. 
RR-s are very slowly relaxing domains, remaining in the opposite phase than
the whole system for a long time, causing slow evolution of the order parameter.
In the entire GP, which is an extended control parameter region around the
critical point, susceptibility diverges and auto-correlations exhibit fat 
tailed, power-law behavior, resulting in bursty behavior~\cite{burstcikk}, 
frequently observed in nature~\cite{Karsai_2018}.
Even in infinite dimensional systems, where mean-field behavior is expected, 
Griffiths effects can occur in finite time windows~\cite{Cota2016}. 

Heterogeneity effects are very common in nature and result in dynamical 
criticality in extended GP-s, in case of quasi-static quenched disorder 
approximation~\cite{Munoz2010}. 
This leads to avalanche size and time distributions, with non-universal 
power-law tails. It has been shown within the framework of modular 
networks~\cite{Munoz2010,HMNcikk,Cota_2018} and a large human connectome 
graph~\cite{gastner_topology_2016,CCdyncikk,CCdyncikk,CCrev}.
The word connectome is defined as the structural network of neural
connections in the brain~\cite{sporns_human_2005}.
Recently the hemibrain has been derived from a 3D image of roughly half 
the fruit-fly (FF) brain.
It contains verified connectivity between ~25,000 neurons that form
more than twenty-million connections~\cite{FlybrainAtl,Xu2020.01.21.911859}.
However, as this is not a complete central nervous system many of the
connections do not connect to the nodes published.

As individual neurons in-vitro emit periodic signals~\cite{PSM16},
it is tempting to use oscillator models and to investigate criticality
at the synchronization transition point. Note, however that according to
other experiments they can also show a variety of spiking behaviors.
Recently, a brain model analysis
using Ginzburg--Landau type equations concluded that empirically reported
scale-invariant avalanches can possibly arise if the cortex is operated at
the edge of a synchronization phase transition, where neuronal avalanches
and incipient oscillations coexist \cite{MunPNAS}.

One of the most fundamental models showing phase synchronization is the
Kuramoto model of interacting oscillators~\cite{kura} and was used to
study synchronization transition on various synthetic and connectome
graphs available~\cite{Frus,Frus-noise,FrusB,KurCC,KKIdeco}.
Note, that the Kuramoto equation, while neglecting the integration feature 
of spiking activity of neighboring neurons, still provides a fundamental, 
mechanistic model for synchronization transition and criticality.
It also involves the quasi-static assumption, according to which
the time scale of network change is much larger than the time scale
of reaching the steady state of the processes running on it.
That means, it is permissible to focus on determining the critical dynamics on
a snapshot of the connectome, not taking plasticity and learning into account.
There is also uncertainty in the KKI-18
full human brain connectome structure as discussed in \cite{CCrev},
but a recent study claims that diffusion tensor imaging is in good
agreement with ground-truth data from histological tract
tracing~\cite{delettre_comparison_2019}.
 
Because of quenched, purely topological heterogeneity an intermediate 
phase was found between the standard synchronous and asynchronous phases, 
showing "frustrated synchronization", meta-stability and chimera-like 
states~\cite{chimera}.
This complex phase was investigated further in the presence of
noise~\cite{Frus-noise} and on a simplicial complex model of manifolds
with finite and tunable spectral dimension \cite{FrusB} as a simple model
for the brain. 

In case of a representative of large human white matter 
connectomes~\cite{gastner_topology_2016} the $N = 804\,092$ node
KKI-18 network GP-s have been found via measuring the
desynchronization times of local perturbations~\cite{KurCC,KKIdeco}.
Now we extend this kind of investigation via Kuramoto model (KM) 
on the FF connectome. 
The comparison of the synchronization transition results on the
KKI-18 and FF is valid, because for FF we know the full
topology of the neural network and for KKI-18 the unknown,
microscopic details below its $1 \ \rm mm^3$ resolution are not
expected to affect the long-wavelength behavior determining
the critical properties. Our model describes a resting state brain.
External sources, leading to the well known Widom line phenomena 
have recently been studied both by experiments and simulations. 
Quasi-criticality, generated by external excitation, was suggested 
to explain the lack of universality observed in
different experiments~\cite{Fosque-20}.


\section{The topology of the fruit-fly connectome }

\begin{figure}[h]
\includegraphics[height=6.5cm]{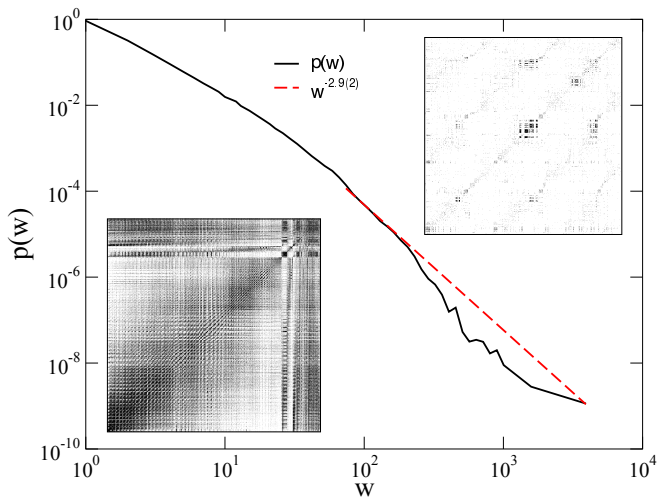}
\caption{Weight distribution of the fruit-fly connectome.
Right inset: adjacency matrix plot of the fruit-fly connectome. 
Left inset: full adjacency matrix down-sampled with a max pooling 
kernel of size $10 \times 10$. 
Black dots denote connections between presynaptic and postsynaptic 
neurons.
Right inset: zoom-in to the center of the matrix without down-sampling.}
\label{indegfig}
\end{figure}

We downloaded the hemibrain data-set (v1.0.1) from~\cite{down-hemibrain1.0.1}.
It has $N_{FF}=21.662$ nodes and $L_{FF}=3.413.160$ edges, out of which
the largest single connected component contains $N=21.615$ and 
$L=3.410.247$ directed and weighted edges, that we used in the simulations.
The number of incoming edges varies between $1$ and $2708$.
The weights are integer numbers, varying between $1$ and $4299$. 
The average node degree is
$\langle k\rangle= 315.129$ (for the in-degrees it is: $157.6$), 
while the average weighted degree is $\langle w\rangle= 628$.
The adjacency matrix is visualized by the insets of Fig.~\ref{indegfig}.
One can see a rather homogeneous, almost structureless network,
however it is not random, as discussed in the graph analysis~\cite{Graphanal}.
For example, the degree distribution is much wider than that of a random 
Erd\H os-R\'enyi (ER) graph and exhibits a fat tail.

The weight distribution $p(w)$ we obtained by exponentially growing bin 
sizes: $w_i \propto 1.12^i$ can can be seen on Fig.~\ref{indegfig}.
Interestingly, the tail of $p(w)$ shows a nontrivial shape, as compared 
to Fig.5 of Ref.~\cite{Graphanal}, where this fine structure cannot be seen, 
due to the linear binning used there. A fitting for the whole weight 
distribution data, assuming a PL with exponential cutoff is published in
~\cite{Graphanal}, which is characterized by the exponent $-1.67$. 
The application of growing bin sizes on the weights of the available traced 
connections does not suggest an exponential cutoff, but a PL tail with 
an exponent $-2.9(2)$ could be fitted for the $w > 100$ region.
We think this might be relevant, because in case of KKI-18 connectome a 
similar PL was found for the tail of link weight distribution and maybe
it is related to an optimal weight distribution (counting the multiplicity
of edges) in real networks embedded in the 3D space.
Of course, due to the partial FF connectome data, we assume that 
the additional, omitted edges result in a tail with a finite 
exponential size cutoff.

The modularity quotient of the FF network defined by
\begin{equation}
Q=\frac{1}{N\av{k}}\sum\limits_{ij}\left(A_{ij}-
\frac{k_ik_j}{N\av{k}}\right)\delta(g_i,g_j),
\end{equation}
is very low: $Q = 0.002264$,
where $A_{ij}$ is the adjacency matrix and $\delta(i,j)$ is the Kronecker
delta function. The weighted modularity quotient is even lower:
$Q_w = 0.0001184$.
In comparison, the modularity quotient of the KKI-18 network is about
$40$ times greater: $Q_w = 0.0047$

The Watts-Strogatz clustering coefficient \cite{WS98} of a
network of $N$ nodes is
\begin{equation}\label{Cws}
C = \frac1N \sum_i 2n_i / [ k_i(k_i-1) ]\ ,
\end{equation}
where $n_i$ denotes the number of direct edges interconnecting the
$k_i$ nearest neighbors of node $i$. This is $C_{FF} = 0.191$,
about $12.36$ times larger than that of a random network of same size:
$C_r=0.01545$, obtained by $C_r = \langle k\rangle / N$.
In case of KKI-18 we found: $C_{KKI-18} = 0.5$.

The average shortest path length is defined as
\begin{equation}
L = \frac{1}{N (N-1)} \sum_{j\ne i} d(i,j) \ ,
\end{equation}
where $d(i,j)$ is the graph distance between vertices $i$ and $j$.
For FF this is $L_{FF} = 2.7531$, about $1.3$ times larger than that
of the random network of same size: $L_r = 2.1162$, following
from the formula~\cite{Fron}:
\begin{equation}
L_r = \frac{\ln(N) - 0.5772}{\ln\langle k\rangle} + 1/2 \ .
\end{equation}

So, the FF is a small-world network, according to the definition of
the coefficient~\cite{humphries_network_2008}:
\begin{equation}
\sigma = \frac{C/C_r}{L/L_r} \ ,
\label{swcoef}
\end{equation}
because $\sigma_{FF} = 9.5$ is much larger than unity.

We estimated the effective graph (topological) dimension,
which is obtained by the breadth-first search algorithm: $d=5.4(5)$, 
which is defined by $N(r) \sim r^d$, where we counted the number of 
nodes $N(r)$ with chemical distance $r$ or less from the seeds and 
calculated averages over the trials. Note, however
that finite size cutoff happens already for $r > 2$.
This dimension renders this model into the mean-field region, because 
the upper-critical dimension is $d_c = 4$.


\section{Numerical analysis of the Kuramoto model}


We used the KM of interacting oscillators~\cite{kura} to study the 
synchronization on the human KKI-18, the FF connectome as well
as on ER random graphs for comparison KM was originally defined on 
full graphs, corresponding to mean-field behavior~\cite{chate_prl}. 
The critical dynamical behavior has recently
been explored on various random graphs~\cite{cmk2016,Kurcikk,KurCC}.
Phase transition in the KM can happen only above the lower critical 
dimension $d_c^-=4$ \cite{HPClett}. In lower dimensions, a true, singular 
phase transition in the $N\to\infty$ limit is not possible, but partial
synchronization can emerge with a smooth crossover if the oscillators are 
strongly coupled.

The KM describes interacting oscillators with phases~$\theta_i(t)$
located at $N$ nodes of a network, which evolve according to the
dynamical equation
\be
\dot{\theta_i}(t) = \omega_{i,0} + K \sum_{j} W_{ij} \sin[ \theta_j(t)- \theta_i(t) ].
\label{kureq}
\ee
Here, $W_{ij}$ is the weighted adjacency matrix and summation is performed over neighboring nodes of $i$. There is a quenched heterogeneity in  $W_{ij}$ as
well as in $\omega_{i,0}$, which is the intrinsic frequency of the $i$-th 
oscillator, drawn from a $g(\omega_{i,0})$ distribution.
The global coupling $K$ is the control parameter of the model by which
we can tune the system between asynchronous and synchronous states.
One usually follows the synchronization transition through studying
the Kuramoto order parameter defined by
\be
R(t)=\frac{1}{N}\left|\sum_{j=1}^Ne^{i\theta_j(t)}\right|,
\label{op}
\ee
which is non-zero above a critical coupling strength $K > K_c$ or tends to
zero for $K < K_c$ as $R \propto\sqrt{1/N}$. At $K_c$, $R$ exhibits growth
as
\be
R(t,N) = N^{-1/2}\, t^{\eta}\, f_{\uparrow}(t / N^{\tilde z}) \ ,
\label{escal}
\ee
with the dynamical exponents ${\tilde z}$ and $\eta$,
if the initial state is incoherent.

Additionally, we have also calculated another order parameter, which
measures the spread of frequencies
\be
\Omega(t,N) = \frac{1}{N} \sum_{j=1}^N (\overline\omega-\omega_j)^2,
\label{Oscal}
\ee
In case of a single peaked self-frequency distribution it is an
appropriate order-parameter, besides the more commonly used measure,
which counts the number of oscillators in the largest
cluster having an identical frequency~\cite{HPCE}.

Generally we used the Runge-Kutta-4 integration algorithm with 
step sizes $\delta=0.01$ or $\delta=0.1$ if it was sufficient, 
via a special, parallel algorithm, running on GPU-s. We have
averaged over the solutions for thousands of different 
initial self-frequencies, chosen randomly from Gaussian distributions
with zero mean and unit variance at each control parameter value.
In a previous paper~\cite{KKIdeco}, we have shown the possibility of 
rescaling these onto more realistic, narrow banded frequencies thanks
to the Galilean invariance of the KM. Some of the runs, especially
for larger couplings $K\ge 3$ , were tested by the adaptive solver 
Bulrisch-Stoer~\cite{BS} of the boost library. For very large couplings,
$K>30$ only the adaptive solver could provide reasonable results.

First we have determined the growth behavior of $R(t)$ of the Kuramoto
equation solution with incoming weight normalization, in order to 
mimic a local homeostasis, provided by the unknown balance of 
inhibition/excitation:
\be
W'_{ij} = W_{ij}/\sum_{j\; \in\; \rm{neighb.\ of} \ i} W_{ij}.
\label{Wnorm}
\ee
This renormalization has been used in previous connectome studies
~\cite{CCdyncikk,KurCC,KKIdeco,Rocha2008,CCrev}.
Recently, a comparison of modeling and experiments arrived at
a similar conclusion: equalized network sensitivity improves the
predictive power of a model at criticality in agreement with
the fMRI correlations~\cite{Rocha2008}.
The solution of equations were started from incoherent states, but for
larger $K$ values it was better to start from coherent states in order
to reach the steady states without large oscillations. 

\begin{figure}[h]
\includegraphics[height=5.5cm]{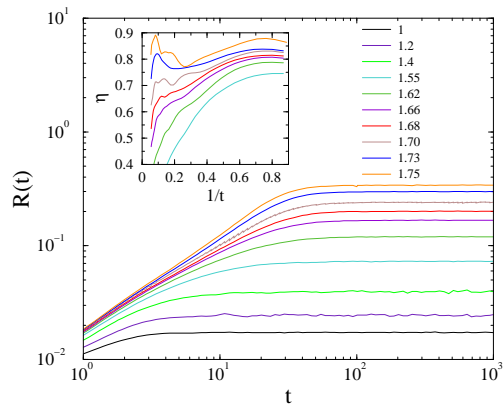}
\caption{Growth of $R(t)$ on the weight normalized FF, using incoherent 
initial state for different $K$ global coupling values as shown by the legends.
One can locate a transition by the convex/concave criterion at $K_c = 1.70(2)$, 
where an initial PL growth can be observed. The inset shows the local slopes 
of the same data on $1/t$ scale with the same color coding
from $K=1.75$ (top curve) to $K=1.55$ (bottom curve).
}
\label{growthfig}
\end{figure}

As Fig.~\ref{growthfig} shows there is a transient region up to $t < 30$
followed by a level-off as the correlation length exceeds the system
size, causing a steady state saturation of the phase synchronization.
In the transient region curves with $K > 1.7$ exhibit an upward, while
those with $K < 1.7$ downward curvature.
To see the corrections to scaling we determined the effective exponents
of $R$ as the discretized, logarithmic derivative of Eq.~(\ref{escal}) at
these discrete time steps $t_k$, near the transition point
\begin{equation}  \label{Reff}
\eta_\mathrm{eff} = \frac {\ln \langle R(t_{k+3})\rangle - \ln \langle R(t_{k})\rangle} 
{\ln(t_{k+3}) - \ln(t_{k})} \ .
\end{equation}
Here the brackets denote sample averaging over different initial conditions.
These effective exponent values can be seen on the local slope inset of 
the figure. Some fluctuation and modulation effects, coming from the
weak modular graph structure of the FF, remain. 
One can estimate a synchronization transition at $K_c = 1.70(2)$, 
characterized by $\eta=0.70(5)$.
This is close to the mean-field value, obtained in~\cite{cmk2016,KurCC}
$\eta_{MF} \simeq 0.75$ and higher than those of the large human white
matter connectomes, where the graph dimension was found to be
$d < d_c = 4$~\cite{gastner_topology_2016}.

\begin{figure}[h]
\includegraphics[height=5.5cm]{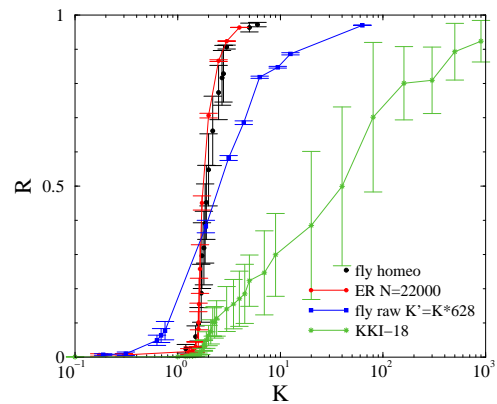}
\caption{Synchronization transition of $R(t\to\infty,K)$ on 
different graphs.}
\label{betafig}
\end{figure}

Using the steady state values we also determined the transition as
the function of the control parameter $K$. Fig.~\ref{betafig} displays
a comparison of the FF transition with the results obtained 
on the KKI-18 human connectome. The transition is sharp around
$K_c = 1.70(2)$ and $R$ changes from $0.02$ to $0.97$ as $K$ from
$1.2$ to $6$.
In comparison, similar change of $R$ for the KKI-18 spans from
$1.6 < K < 10^3$.
We also plotted the results obtained without the application
of weight normalization by running on the raw FF network on 
Fig.\ref{betafig}.
In this case the transition occurs at a much lower coupling: 
$K_c = 0.00090(5)$, so we multiplied them on the plot by the average
weight value $K' = K\times 628$. Note, that the transition of the 
raw case is not smoother than the homeostatic one, just it appears 
to be like that, as the consequence of the linear up-scaling of $K$.
It happens in the $0.0005 < K < 0.2$ region.
The steady state results on a random ER graph with $N=22.000$ and 
$\langle k\rangle= 315$ are also displayed. Here we used 
weight normalization condition (\ref{Wnorm}) as for FF. 
The synchronization transition occurs in the $1.4 < K < 5$ region,
which is slightly narrower than for the FF.

\begin{figure}[h]
\includegraphics[height=5.5cm]{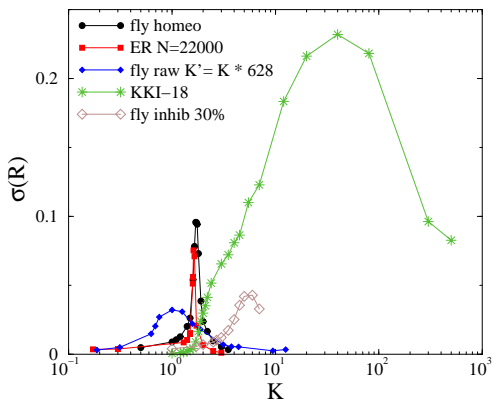}
\caption{Fluctuations of $R(t\to\infty,K)$ for different graphs.}
\label{beta-flucfig}
\end{figure}

We can analyze the transition further by determining the fluctuations 
of $R(t\to\infty,K)$ near the transition. 
This is plotted on Fig.~\ref{beta-flucfig}. As we can see the 
standard deviation: $\sigma ( R(t\to\infty,K) )$ of the FF
is very similar to that of an ER graph of same size and average degree, 
but somewhat wider. In comparison the KKI-18 exhibits a much more smeared 
transition region, even though the weighted average degree
is smaller: $\langle w\rangle_{\mathrm {KKI-18}} = 448$ than that of the fly
connectome: $\langle w\rangle_{\mathrm {FF}}= 628$. 
As $d_{KKI-18} < d_c^-=4$ this is a crossover transition and no
exact finite scaling is applicable to rescale it.

In case of KM on random ER graphs increasing the size causes small 
decrease of $K_c$ as well as narrower peaks as shown in~\cite{KurCC}. 
If we increase the average degree from $\langle k\rangle = 4$ to 
$\langle k\rangle = 350$, the critical point $K_c \simeq 0.482$ 
moves to $K_c \simeq 1.65$ close to that of the full graph case 
$K_c \simeq 1.6$. Thus, one may expect that the bigger average 
degree of FF would cause a peak at larger couplings.
In contrary we can see that the Hierarchical Modular Network (HMN) structure 
of KKI-18 causes nontrivial effects on the $\sigma(R)$ peak and on the 
width of the phase synchronization transition region.

We have also investigated the frequency synchronization order parameter,
which is defined here as Eq.(~\ref{Oscal}). 
In case of the single peaked Gaussian self-frequencies one can follow the 
frequency entrainment by this quantity. This has the advantage of having 
lower critical dimension: $d_c^- = 2$ as compared to the phases: $d_c^- = 4$.
This was showed on regular lattices~\cite{HPClett}, but Ref.~\cite{Biatopdim}
obtained similar conclusion on complex network.  
Thus in case of graph, like the KKI-18, a real frequency phase-transition
can occur, if we found the the human brain to exhibit topological dimension
$d > 2$, even for higher resolutions.

\begin{figure}[h]
\includegraphics[height=5.5cm]{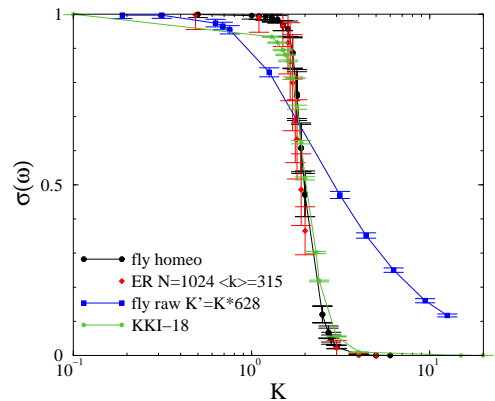}
\caption{Frequency entrainment order parameter $\Omega(t\to\infty,K)$
of the KM obtained on different graphs.}
\label{omegasfig}
\end{figure}

\begin{figure}[h]
\includegraphics[height=5.5cm]{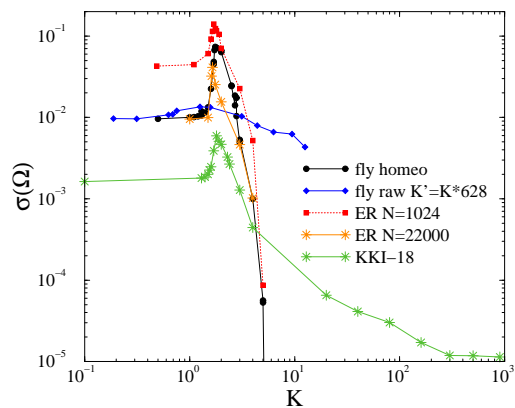}
\caption{Fluctuations of the frequency entrainment order parameter
$\sigma (\Omega (t\to\infty,K) )$ for different graphs.}
\label{omega-flucfig}
\end{figure}

Indeed as the Fig.~\ref{omegasfig} shows the frequency transitions
on the fly on the ER and on the human KKI-18 are very similar.
Now the finite size scaling
\be
| K - K_c | \propto N^{-1/\tilde\nu}
\ee
is applicable as all of these graphs have $d > d_c^-$.
By considering the fluctuations of this order parameter:
$\sigma (\Omega (t\to\infty,K) )$ we find that the peaks are close,
but the KKI-18 transition region is much wider in the high
coupling region, than in case of the ER and the FF  (see ~\ref{omega-flucfig}).
The fluctuation region on the random ER graph is the narrowest
and the peak value decreases as we increase $N$.
We have also plotted the results obtained on the raw FF graph,
up-scaled by the average value of the weights: $K'=K\times 628$.
The distribution looks wider, but this is just the consequence of the
horizontal rescaling.

Finally, we also performed measurements for the desynchronization 
times as in \cite{KurCC,KKIdeco}.
To define desynchronization "avalanches" in terms of the Kuramoto order parameter,
we can consider processes, starting from fully de-synchronized initial states by
a single phase perturbation (or by an external phase shift at a node), followed
by growth and return to $R(t_x) = 1/\sqrt{N}$, corresponding to the disordered
state of $N$ oscillators. In the simulations one can measure the first return,
crossing times $t_x$ in many random realizations of the system.
In ~\cite{KurCC,KKIdeco}, the return or spontaneous desynchronization
time was estimated by $t_x = (t_k + t_{k-1})/2$,
where $t_k$ was the first measured crossing time, when $R(t_k)$ fell below
$1/\sqrt{N}$.

\begin{figure}[h]
\includegraphics[height=5.5cm]{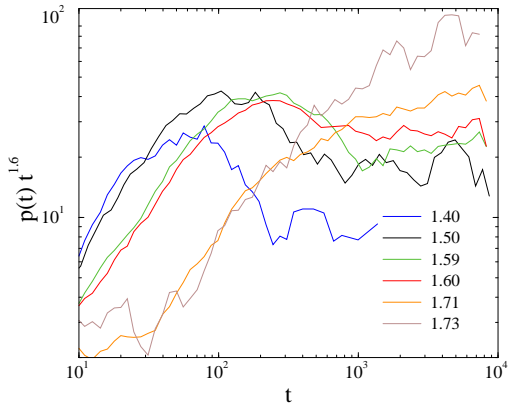}
\caption{Desynchronization time distributions $p_x(t) t^{1.6}$
near the transition point in case of the fly connectome with local
homeostasis for different couplings, as shown by the legends.}
\label{elo-ThrO.fly}
\end{figure}

Following a histogramming procedure, one can obtain $p_x(t)$ distributions
shown on Fig.\ref{elo-ThrO.fly} for the weight normalized, homeostatic case.
For $K = 1.65(5)$ (i.e.\ near the transition point estimated before),
we can find critical PL decay characterized by $\tau_t\simeq 1.6(1)$, close
to the mean-field value of the spontaneous desynchronization of $R(t)$, 
as defined in~\cite{KurCC}.
For $K > 1.7$ the curves decay as $p_x(t) \sim 1/t$ up to a cutoff,
corresponding to the ordered state, while for $K < 1.5$ the curves
break down sharply. It is hard to decide if there is a narrow GP
in the $1.5 \le K \le 1.7$ region due to the strong fluctuations
remained even after averaging over tens of thousands of samples with
different $\omega_{i,0}$ initial conditions.

\begin{figure}[h]
\includegraphics[height=5.5cm]{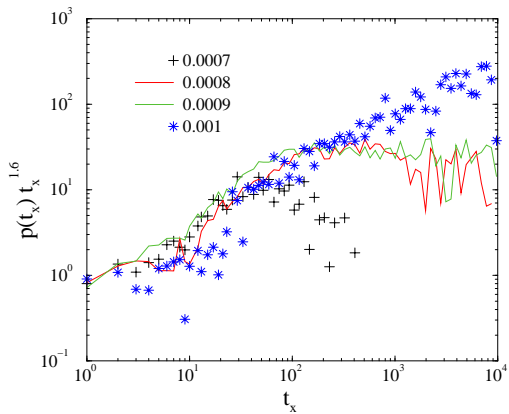}
\caption{Desynchronization time distributions $p_x(t) t^{1.6}$ near
the transition point, using the raw fly connectome graph for
different couplings, as shown by the legends.}
\label{elo-ThrFrn}
\end{figure}

Similar results have been obtained using the raw FF graph,
as shown on Fig.\ref{elo-ThrFrn}. The transition point is at
$K_c=0.0008(1)$, where we can observe a saturation of the
$p_x(t) t^{1.6}$ for $t > 200$, thus again mean-field scaling occurs.
At $K=0.001$ we can also see the $p_x(t) \propto 1/t$ decay, corresponding
to the synchronized state, in which arbitrarily large decay times can
happen, but no signs of sub-critical PL-s, corresponding to a GP have been
found.

\begin{figure}[h]
\includegraphics[height=5.5cm]{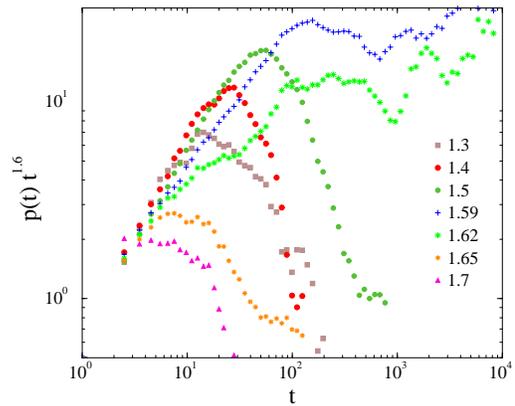}
\caption{Desynchronization time distributions $p_x(t) t^{1.6}$ near
the transition point in case of ER graphs with different couplings,
as shown by the legends.}
\label{elo-ER}
\end{figure}

For comparison we have done this analysis for full and for ER graphs
with $N=22.000$ and $\langle k\rangle = 315$.
Now we just show the results for the ER case on Fig.\ref{elo-ER}.
Below $K_c \simeq 1.59$ the $p_x(t) t^{1.6}$ curves break down quickly,
without any sign of PL tails. While for $K=1.59$ we see a saturation
for $t > 200$, the  $K=1.62$ curve seems to cross over to the
singular $p_x(t) \sim 1/t$ behavior. Going beyond this the curves
break down very quickly again, suggesting that within the maximum
measurement time $t=10^4$ desynchronization events could not happen.

We have also tested the effects of the introduction of negative
couplings by flipping the sign of outgoing weight values:
$W_{ij} = -W_{ij}$ of $30\%$ of randomly chosen nodes.
As the consequence the transition region broadens considerably as
shown on Fig.\ref{beta-flucfig}.


\section{Conclusions}


In conclusion we have investigated KM at the edge of synchronization
by comparing the dynamical behavior on the FF, ER and the KKI-18 large 
human connectome. The FF network topology is rather similar, almost 
free of modules, to the high-dimensional ER graph. Thus we found 
a mean-field like behavior, unlike for the KKI-18, which has 
$d < 4$, a HMN structure, which enhances and broadens the transition 
region with the appearance of GP singularity.
Although the link weight distribution of FF exhibits a fat tail, 
it does not seem to be enough to introduce visible GP effects, 
or maybe a very weak ones. Thus one can think that the fly brain's 
simpler structure does not allow the appearance of the complex sub-critical
dynamical phenomena, which are present in the human brain.
The lack of modules on smaller scales may also explain that in global human
brain measurements non- universal scaling~\cite{brainexp} are reported, while
local electrode studies~\cite{BP03} show mean-field exponents.
Possibly electrode studies~\cite{BP03} measure local
activity and within those small volumes modules and GP are less relevant
than on the whole brain scale. 

The range of the synchronization transition region is slightly broader
than in case of the ER, but much narrower than in case of the KKI-18
when we applied link weight normalization, to mimic local homeostasis.
This is shown both by the phase and frequency order parameters.
Without link weight normalization the KM transition occurs at very
low coupling values, but shows mean-field scaling. This was shown by
measuring the synchronization growth exponent $\eta$ and
the desynchronization time exponent $\tau_t$.

If we allow negative couplings the transition region broadens further,
leading to a spin glass like phase, where GP effects may also emerge.
But as the details and dynamics of such negative couplings are unknown
in case of the FF-s we have not investigated this further.
We have arrived to similar conclusions as the very recent publication
by Buendia et al~\cite{buendia2021broad} in case of the complex interplay
between structure and dynamics, but we showed the emergence of a
critical transition in terms of desynchronization times as well as
the initial-slip, characterized by the exponent $\eta$.

Given the limitations and assumptions we mentioned in the   
Introduction we have provided ample numerical evidence for the
different dynamical critical behavior of the Kuramoto model,
as the result of the different connectome topology of a fly 
and of a human brain.
Further studies on other animals, preferably mammals should be 
performed in order to fully justify the proposition expressed in the title.

\section*{Acknowledgments}
  
We thank R\'obert Juh\'asz and Shengfeng Deng for useful comments and 
discussions.
G.\'O. is supported by the National Research, Development and Innovation
Office NKFIH under Grant No. K128989 and the Project HPC-EUROPA3
(INFRAIA-2016-1-730897) from the EC Research Innovation Action under the
H2020 Programme. We thank access to the Hungarian national supercomputer 
network NIIF and to BSC Barcelona.

\bibliography{main}

\end{document}